\documentclass{PoS}
\usepackage[utf8]{inputenc} 
\usepackage{booktabs,amsmath}
\usepackage{bm,amsmath,amssymb,amsfonts,latexsym,psfrag}
\usepackage{amssymb}
 \usepackage{array}
\usepackage{mathtools}
\usepackage{amsthm}
\usepackage{newlfont}
\usepackage{esdiff}
\usepackage{multirow}
\usepackage{graphics}
\usepackage{graphicx}
\usepackage{color}
\usepackage{tabularx}
\usepackage{bm}
\title{Perturbative unitarity bounds for fermions composite models}

\ShortTitle{Perturbative unitarity bounds for fermions composite models}

\author{\speaker{S. Biondini}\thanks{The speaker thanks R.~Leonardi, O.~Panella and M.~Presilla for collaboration on the work \cite{Biondini:2019tcc} presented at the workshop. }\\
        Van Swinderen Institute, University of Groningen, 9747 AG Groningen,
        \\
The Netherlands\\
        E-mail: \email{s.biondini@rug.nl}}

\abstract{Perturbative unitarity is a powerful tool for inferring the range of validity of a given effective field theory. Here, we study such a bound in the parameter space of dimension-5 and dimension-6 effective operators that arise in a scenario of fermion compositeness. These operators are routinely used in experimental searches at the LHC to constraint contact and gauge interactions between ordinary Standard Model fermions and excited states of mass $M$.
We derive the unitarity bound for the production process of an excited neutrino, then we implement such bound   
and compare it with the recent experimental exclusion curves for Run 2, the High-Luminosity and High-Energy configurations of the LHC. The results also apply to the searches where a generic single excited state is produced via dimension-6 contact interactions.
The unitarity bound, so far overlooked in these effective models, is quite compelling and can serve as a guide for exploring the parameter space ($M,\Lambda$) in addition to the standard request $M \le \Lambda$. }

\FullConference{Corfu Summer Institute 2019 "School and Workshops on Elementary Particle Physics and Gravity" (CORFU2019)\\
		31 August - 25 September 2019\\
		Corfù, Greece}

\begin{document}

\section{Motivation and introduction}
It is well known that partial-wave unitarity is a powerful tool to estimate the perturbative validity of effective field theories (EFTs). It has been used to provide useful insights both in strong and electroweak interactions~\cite{Lee:1977yc}. Perhaps, the best known example is the  bound on the Higgs mass derived from an analysis of $WW\to WW$ scattering within the Standard Model (SM)~\cite{Lee:1977yc,Lee:1977eg}. On the other end, unitairty has also been applied to a number of approaches  beyond the Standard Model (BSM). For instance in composite Higgs models~\cite{DeCurtis:2016scv}, in searches of scalar di-boson resonances~\cite{DiLuzio:2017qgm,Luzio2017xxx}, searches for dark matter effective interactions~\cite{Endo:2014mja} and on generic dimension-6 operators~\cite{Corbett:2017qgl}.   
In the latter case, effective higher dimensional operators appear and the unitarity bound is provided for different operators structures. 

Effective operators
also enter in composite-fermions scenario which offers a possible solution to the hierarchy pattern of fermion masses~\cite{Pati:1975md,Pati:1983dh,Harari:1982xy,Greenberg:1974qb,Dirac:1963aa} and routinely pursued in high-energy experiments in their phenomenological setting.
In this context~\cite{Terazawa:1976xx,Eichten:1979ah,Eichten:1983hw,Cabibbo:1983bk,Baur:1989kv,Baur:1987ga}, SM quarks ``$q$'' and leptons ``$\ell$'' are assumed to be bound states of some as yet not observed fundamental constituents generically referred 
as {\it preons}. If quarks and leptons have an internal substructure, they are expected to be accompanied by heavy excited states $\ell^*, q^*$ of masses $M$ 
that should manifest themselves at an unknown energy scale, that we label as the compositeness scale 
$\Lambda$.

As customary in an EFT approach, the effects of the high-energy physics scale, here the compositeness scale $\Lambda$, are captured in higher dimensional operators that describe processes at a lower-energy domain, where the fundamental building blocks of the theory are not manifest. The heavy excited states may interact with the SM ordinary fermions via dimension-5 gauge interactions of the SU(2)$_L \otimes$ U(1)$_Y$ SM gauge group of the magnetic-moment type. This way the electromagnetic current conservation is not spoiled by e.g.~$\ell^* \ell \gamma$ processes~\cite{Cabibbo:1983bk}. In addition, the exchange of preons and/or 
binding quanta of the unknown interactions between ordinary fermions ($f$) and/or the excited states ($f^*$) results in effective
contact interactions (CI)~\cite{Baur:1989kv,Baur:1987ga,Peskin:1985symp,Tanabashi:2018oca}. 
%Contact interactions between ordinary and/or excited fermions may arise by constituent exchange, if the fermions have common constituents, and/or  by exchange of the binding quanta of the new unknown interaction whenever such binding quanta couple to the constituents of both particles~\cite{Peskin:1985symp,Tanabashi:2018oca}. 
In the latter case, the dominant effect is expected to be given by the dimension-6 four-fermion  interactions  scaling with the inverse square of the compositeness scale $\Lambda$:
\begin{equation}
\mathcal{L}^{6}=\frac{g_\ast^2}{\Lambda^2}\frac{1}{2}j^\mu j_\mu \, ,
\label{Lcontact}
\end{equation}
where the current reads 
\begin{equation}
j_\mu=\eta_L\bar{f_L}\gamma_\mu f_L+\eta\prime_L\bar{f^\ast_L}\gamma_\mu f^\ast_L+\eta\prime\prime_L\bar{f^\ast_L}\gamma_\mu f_L + h.c. +(L\rightarrow R) \, ,
\label{Jcontact}
\end{equation}
with $g_*^2 = 4\pi$ and the $\eta$'s factors are usually set equal to unity. As customary in the phenomenological studies, the right-handed currents will be neglected for simplicity. We notice that this is also the setting adopted by the experimental ATLAS and CMS collaborations.

Gauge interactions (GI) were actually conceived first. Here, the excited fermions are collected in weak isospin doublets in the original formulation, whereas the so-called sequential and mirror model were also considered for the excited neutrinos\cite{Takasugi:1995bb,Olive_2014}. We consider here the mirror case,  
where the excited neutrino and the excited electron are grouped into left-handed singlets and  a right-handed SU(2) doublet:
$
e^*_L,  \nu_L^*\, , 
L^*_R=\left(
 \nu^*_R \, , e^*_R 
\right)^T
$,
so that a magnetic-type coupling between the left-handed SM doublet and the right-handed excited doublet via the SU(2)$_L \otimes$ U(1)$_Y$ gauge fields can be written down~\cite{Cabibbo:1983bk,Takasugi:1995bb}:
\begin{equation}
\label{mirror}
{\cal L}^5 =\frac{1}{2 \Lambda} \, \bar{L}_R^* \sigma^{\mu\nu}\left( gf \frac{\bm{\tau}}{2}\cdot \bm{W}_{\mu\nu} +g'f' Y B_{\mu\nu}\right) L_L +h.c. \, .
\end{equation}
Here, $L^T =({\nu_\ell}_L, \ell_L)$ is the ordinary lepton doublet, $g$ and $g'$ are the SU(2)$_L$ and U(1)$_Y$ gauge couplings and $\bm{W}_{\mu\nu}$, $B_{\mu\nu}$ are the field strength tensor of the corresponding gauge fields respectively; $\bm{\tau}$ are the Pauli matrices and $Y$ is the hypercharge,  $f$ and $f'$ are dimensionless couplings and are expected (and assumed) to be of order unity. 
\begin{figure}[t!]
\centering
\includegraphics[scale=0.33]{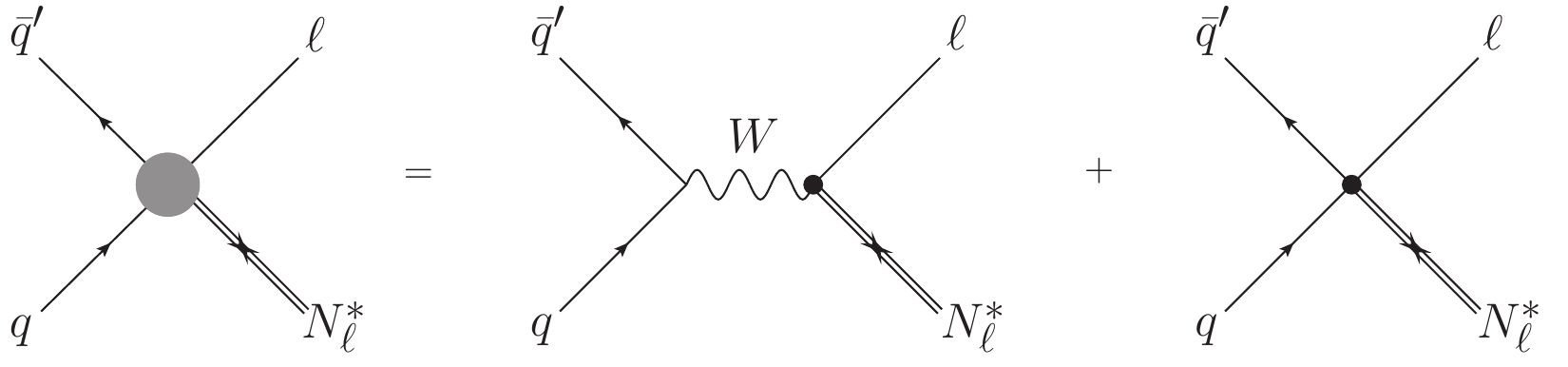}
\caption{\label{fig:process}Feynman diagrams depicting the mechanisms responsible for the process $q \bar{q}' \to N^* \ell$, where $\ell$ stands for both $\ell^\pm$. The dark grey blob (diagram on the left) describes the production of an on-shell heavy Majorana neutrino $N$ in proton-proton collisions at
LHC. The production is possible both with gauge interactions (first diagram on the right-hand side) and with four-fermion contact
interactions (second diagram on the right-hand side).}
\end{figure}

Excited states interacting with the SM sector through the model Lagrangians (\ref{Lcontact})-(\ref{Jcontact}) and (\ref{mirror}) have been extensively searched for at high-energy collider facilities and their phenomenology has also been discussed in a series of recent papers~\cite{Biondini:2012ny,Majhi:2012bx,Leonardi:2014epa,Biondini:2014dfa,Leonardi:2015qna,Panella:2017spx,Guo:2016hjt,Presilla:2018ryu,Caliskan:2018vsk,Ozansoy:2019kdq}. The current  strongest bounds are due to the recent LHC experiments. Charged leptons ($e^*
,\mu^*$) have been searched for in the channel $pp\to \ell \ell^* \to \ell \ell \gamma$~\cite{atlas-limit,atlas-limit-new,cms-limit-7TeV,cms-limit-8TeV,cms-limit-13TeV,CMS-PAS-EXO-18-004,Sirunyan:2018zzr,Sirunyan:2020awe}, i.e. produced via CI and then decay via GI, and in the channel $pp \to \ell \ell^* \to \ell \ell q\bar{q}^\prime$  \cite{CMS-PAS-EXO-18-013} where both production and decay proceed through CI.
Neutral excited leptons have been also discussed in the literature and the corresponding phenomenology at LHC has been discussed in detail in the case of a heavy composite Majorana neutrino ~\cite{Leonardi:2015qna}, which we label generically with $N^*$ in the following. 
A dedicated experimental analysis has been carried out 
by the CMS collaboration \cite{Sirunyan:2017xnz} on LHC data collected for $\sqrt{s}$ = 13 TeV and looking for the process \begin{equation}pp\rightarrow \ell N^*_\ell \,  \rightarrow \ell\ell q\bar{q}^\prime\end{equation}
with dilepton (dielectrons or dimuons) plus diquark final states.  The particle $N^*_\ell$ is excluded for masses up to $4.60$ $(4.70)$ TeV at $95\%$ confidence level, assuming $M = \Lambda$. Moreover, possible connections between composite Majorana neutrinos of this model and baryogenesis via leptogenesis have been explored in refs.~\cite{Zhuridov:2016xls,Biondini:2017fut}.

We emphasize that in all phenomenological studies referenced above, as well as all experimental analyses that have searched for excited states at colliders, it is customary to impose the constraint $M \le \Lambda$ on the parameter space of the model.  Unitarity bounds have never been taken into account or discussed in connection with the effective interactions of the fermion excited states.
In this contribution, we report that instead the unitarity bounds, as extracted from Eq.~(\ref{Lcontact})-(\ref{Jcontact}) and (\ref{mirror}), are quite compelling and should be included in future analyses based on such effective composite models because they constraint rather strongly the parameter space.  
While we present an explicit calculation of the unitarity bound for heavy composite neutrino searches, we expect that similar compelling bounds would apply for excited electrons ($e^*$), muons ($\mu^*$) and quarks ($q^*$). Indeed, the effective operators that describe the latter excited states have the very same structure of those applicable to the composite neutrinos. This is especially true for the contact interactions operators.   
\section{Derivation of the perturbative unitarity bound and implementation}
The unitarity bound can be derived by standard methods that makes use of the optical theorem and the expansion of the scattering amplitude in partial waves. In order to specify the CI and GI Lagrangians for a definite situation, we consider here the production of the excited Majorana neutrino at the LHC. 

The key object is the interacting part of the $S$ matrix, indicated with $T$, that enters the partial-wave decomposition of the scattering amplitude as follows 
\begin{equation}
\mathcal{M}_{i \to f} (\theta ) = 8 \pi \sum_j (2 j +1 ) T^j_{i \to f} d^j_{\lambda_f \, \lambda_i} (\theta) \, ,
\label{PW_dec}
\end{equation}
where $j$ is the eigenvalue of the total angular momentum $J$ of the incoming (outgoing) pair,  $d^j_{\lambda_f \, \lambda_i} (\theta)$ is the Wigener d-function and $\lambda_i$ ($\lambda_f$) is the total helicity of the initial (final) state pair. We consider azimuthally symmetric processes and fix $\phi=0$ accordingly.  From the optical theorem and the decomposition in Eq.~(\ref{PW_dec}), one can find the perturabtive unitarity condition of an inelastic process to be
\begin{equation}
\sum_{f \neq i} \beta_i \beta_f |T^j_{i \to f}|^2 \leq 1 \, ,
\label{Bound_Inelastic}
\end{equation}
that holds for each $j$, and  where $\beta_i$ ($\beta_f$) is the factor of the two-body phase space and reads for two generic particles with masses $m_1$ and $m_2$
\begin{equation}
\beta = \frac{\sqrt{\left[ \hat{s}-(m_1-m_2)^2\right] \left[ \hat{s}-(m_1+m_2)^2\right]}}{\hat{s}} \, .
\label{Beta_fi}
\end{equation}
For  $m_1=m_2$ one recovers the particle velocity. The unitarity bound is imposed on the subprocess involving the proton valence quarks, namely $q \bar{q}' \to \ell N^*_\ell$ as shown in Figure~\ref{fig:process}. Then, the relevant interaction(s) read
    \begin{equation}
    \mathcal{L}_{\text{CI}}= \frac{g_*^2 \, \eta}{\Lambda^2} \bar{q}' \gamma^\mu P_L q \, \bar{N} \gamma_\mu P_L \ell  + h.c. \, ,
    \label{CI_Lag_LH}
\end{equation}

\begin{equation}
    \mathcal{L}_{\text{GI}}= \frac{g \, f}{\sqrt{2} \Lambda} \bar{N} \sigma^{\mu \nu} (\partial_\mu W_\nu^+) P_L \ell + h.c. \, .
    \label{GI_Lag_LH}
\end{equation}
Accordingly, in Eq.~(\ref{Beta_fi}), $\hat{s}$ denotes the center-of-mass energy in each collision and it is obtained from the nominal collider energy and the parton momentum fractions as $\hat{s}= x_1 x_2 \, s$.  As far as the kinematic is concerned, we set $\beta_i=1$ since the valence quark masses are negligible with respect to the center-of-mass energy at the LHC. On the other hand, one finds  
$\beta_f = 1-M^{2}/\hat{s}$ for the final state, where the composite neutrino mass has to be kept. 

The core of the method relies on the derivation of the amplitude for the process of interest induced by the contact and gauge-mediated effective Lagrangians (\ref{CI_Lag_LH}) and (\ref{GI_Lag_LH}). Then, one matches the so-obtained result for $\mathcal{M}_{i \to f}$ with the r.h.s of Eq.~(\ref{PW_dec}) and extracts the corresponding $T^j_{i \to f}$ for each definite eigenvalue of the total angular momentum ($j$). The latter are inserted into Eq.~(\ref{Bound_Inelastic}) in order to derive the unitarity condition that the model parameters ($\Lambda, M, g_*,g$) and the center-of-mass energy have to satisfy.  The amplitude $\mathcal{M}_{i \to f}$ is decomposed in terms of definite helicity states. We express the initial and final state particles spinors accordingly \cite{Jacob:1959at}. The helicity of each particle in the initial or final state is $\lambda=\pm 1/2$, being all the involved particles fermions. We label with $\pm$ the initial and final state helicity combinations, namely $(+,+)$, $(+,-)$, $(-,+)$ and $(-,-)$. Since the incoming and outgoing particles travel in opposite directions, the helicities in the Wigner d-functions are defined as $\lambda_i=\lambda_q - \lambda_{\bar{q}'}$ and $\lambda_f=\lambda_{N^*} - \lambda_{\ell}$.  One can adopt two different bases for expressing the spinors and the gamma matrices, the Dirac and chiaral bases (see e.g. appendix in ref.~\cite{Endo:2014mja}). %We used both the options to derive $T^j_{i \to f}$ and checked that our findings are indeed  invariant upon the choice of the basis.

We give the result for the CI Lagrangian in Eq.~(\ref{CI_Lag_LH}) first. The non-vanishing helicity amplitudes read  %\begin{eqnarray}
 %   \mathcal{L}_{\hbox{\tiny CI}}&=& \frac{g_*^2}{\Lambda^2} \bar{q'} \gamma^\mu P_L q \, \bar{N} \gamma_\mu P_L \ell \, , 
 %   \label{dec_L}
%\end{eqnarray}
%that are 
    \begin{eqnarray}
    && T^{j=1}_{(-,+) \to (-,+)}  = -\frac{\hat{s} \, g_*^2}{12 \pi \Lambda^2} \left( 1-\frac{M^{2}}{\hat{s}} \right)^{\frac{1}{2}} \, ,
\label{CI_Tampl_1}
    \\
    && T^{j=1}_{(-,+) \to (+,+)} =   \frac{\sqrt{\hat{s}} \, M \, g_*^2}{12 \sqrt{2}\pi \Lambda^2} \left( 1-\frac{M^{2}}{\hat{s}} \right)^{\frac{1}{2}} \, .
\label{CI_Tampl_2}
\end{eqnarray}
Only the amplitude with $j=1$ is non-zero, due to the initial helicity state. The same occurs with the vector and axial-vector operators studied in \cite{Endo:2014mja} for dark matter pair production at colliders. We notice that a finite composite neutrino mass allows for the helicity flip in the final state originating the term in Eq.~(\ref{CI_Tampl_2}). We obtain the same result if we work with right-handed particles in the CI operator, however the helicities in Eqs.~(\ref{CI_Tampl_1}) and (\ref{CI_Tampl_2}) flip as $+ \leftrightarrow -$.  Using Eq.~(\ref{PW_dec}) and summing over the non-vanishing final helicity states, we obtain
\begin{eqnarray}
    \frac{g_*^4 \, \hat{s} \, (2\hat{s}+M^{2})}{288 \pi^2 \Lambda^4} \left( 1-\frac{M^{2}}{\hat{s}} \right)^{2} \leq 1 \, .
    \label{UNI_CI}
\end{eqnarray}

As far as the GI process is concerned, we proceed the same way. A dimension-5 operator is involved and, in this case, the $W$ boson mediates the scattering between the initial and final states. We keep the $W$ boson mass in our expression and here the SM electroweak current enters besides the one from the composite model. The helicity amplitudes are found to be
\begin{eqnarray}
    && T^{j=1}_{(-,+) \to (-,+)}  = -\frac{i g^2}{24 \pi \Lambda} \frac{\hat{s}^{3/2}}{\hat{s}-m_W^2}\left( 1-\frac{M^{2}}{s} \right)^{\frac{1}{2}} \, ,
    \\
    && T^{j=1}_{(-,+) \to (+,+)} =   \frac{i g^2}{24 \sqrt{2}\pi \Lambda} \frac{\hat{s} \, M}{\hat{s}-m_W^2} \left( 1-\frac{M^{2}}{\hat{s}} \right)^{\frac{1}{2}} \, ,
    \end{eqnarray}
    and the corresponding result for the unitarity bound is 
    \begin{eqnarray}
    \frac{g^4 }{1152 \, \pi^2 \Lambda^2} \frac{\hat{s}^2 \, (2\hat{s}+M^{2})}{(\hat{s}-m_W^2)^2}  \left( 1-\frac{M^{2}}{\hat{s}} \right)^{2} \leq 1 \, .
    \label{UNI_GI}
\end{eqnarray}
A comment is in order. The unitarity bound in Eq.~(\ref{UNI_CI}) is valid for the more generic production process $q \bar{q}' \to f^* f$, i.e. excited charged or neutral leptons and excited quarks accompanied by a SM fermion. This statement traces back to the particle-blind choice adopted in the CIs framework, where the $\eta$'s are set to unity in all the cases. More care has to be taken about a wider applicability of the result for GIs in Eq.~(\ref{UNI_GI}). Here, different factors can enter according to the gauge couplings and gauge bosons that describe the processes involving excited charged leptons and quarks instead of composite neutrinos.  

Next, the  perturbative unitarity bounds are applied to experimental searches in the dilepton and a large-radius jet channel with the CMS detector for the three different collider scenarios \cite{Sirunyan:2017xnz,CMS-PAS-FTR-18-006,Sirunyan:2020awe}.
As already clear from the rather
different coupling values entering the Lagrangians (\ref{CI_Lag_LH}) and (\ref{GI_Lag_LH}), namely $g/ \sqrt{2} \approx 0.4$ versus $g_*^2=4\pi$,  the production mechanism of a heavy composite neutrino and other excited states is dominated by the contact interaction mechanism~\cite{Panella:2017spx}. In particular, it was shown that cross sections in contact-mediated production are usually more than two orders of magnitude larger than the gauge mediated ones for all values of the $\Lambda$ and $M$ relevant in the analyses. This means that it is a reasonable approximation to consider only the bounds given in Eq.~(\ref{UNI_CI}) to constraint the unitarity violation of the signal samples.
%the number of generated signal events with gauge-mediated interactions is

We need to implement the bounds  in the case of hadron collisions at the LHC and, therefore, the square of the centre-of-mass energy of the colliding partons system $\hat{s}=x_1 x_2 s$ is needed. Here, $x_1$ and $x_2$ are the parton momentum fractions and $\sqrt{s}$ is nominal energy of the colliding protons. Since $\hat{s}$ does not have a definite value, we have estimated $\hat{s}$ in each event generated in the Monte Carlo (MC) samples, and we have plugged the result into Eq.~(\ref{UNI_CI}) in order to obtain level curves on the parameter space for which the unitarity bound is satisfied to some degree.
A violation of such bound would signal the breakdown of the EFT expansion and call for higher order operators in $\sqrt{\hat{s}}/ \Lambda$ to help in restoring the unitarity of the process. Therefore, we implement a theoretical uncertainty by allowing up to 50\% of the events to violate the bound. The MC samples for the signal are generated at tree-level with CalcHEP (v3.6)~\cite{Belyaev:2012qa} for $\sqrt{s}=13, 14$ and $27$ TeV proton-proton collisions, with the NNPDF3.0 parton distribution functions \cite{Ball:2014uwa}. 
The information on the parton momenta is then retrieved from the Les Houches Event (LHE) files of each signal process through MadAnalysis~\cite{Conte:2012fm} in order to obtain $\hat{s}$ in each simulated event.

\section{Application to current and future searches of excited fermions}
\begin{figure}[ht!]
\includegraphics[width=0.48\textwidth]{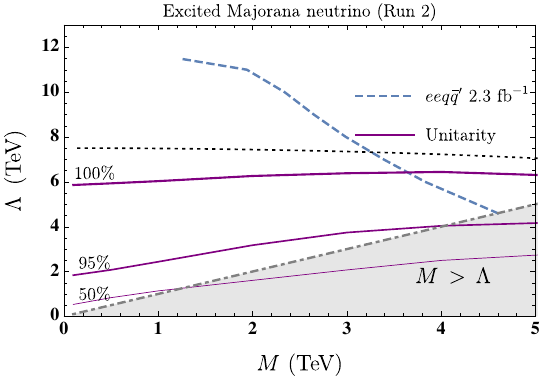}  
\includegraphics[width=0.48\textwidth]{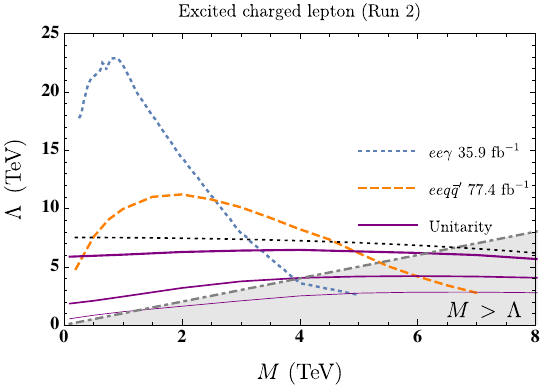}  
\caption{The unitarity bound in the ($M,\Lambda$) plane compared with the Run 2 exclusion at 95\% CL from \cite{Sirunyan:2017xnz}, dashed line (blue), for the $eeq\bar{q}'$ final state signature. The solid (violet) lines with decreasing thickness represent the unitarity bound respectively for 100\%, 95\% and 50\% event fraction satisfying Eq.~\ref{UNI_CI}. The dot-dashed (gray) line stands for the $M= \Lambda$ condition. Here and in the following figures both $\Lambda$ and $M$ start at 100 GeV, and the dotted (black) curve corresponds to the theoretical unitarity bound (Eq.~\ref{UNI_CI} with $\hat{s}=s$). The plot on the left, we show the compaison with the Run 2 for charged leptons searches with two different final states~\cite{CMS-PAS-EXO-18-013}.}
\label{figex1} 
\end{figure}

\begin{figure}[t!]
\includegraphics[width=0.48\textwidth]{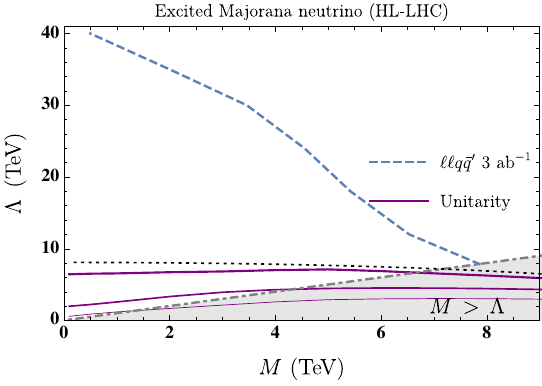}  
\includegraphics[width=0.48\textwidth]{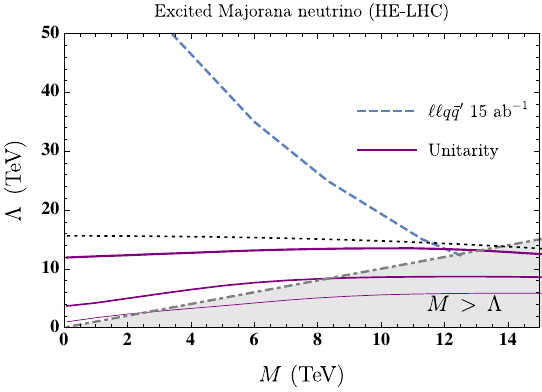}  
\caption{ (Left) The unitarity bound in the plane $(M, \Lambda) $ for the three event fractions as in Fig.~\ref{figex1} compared with the exclusion from the High Luminosity projections study in \cite{CidVidal:2018eel} for LHC at $\sqrt{s}=14$ TeV at 3 ab$^{-1}$ of integrated luminosity and (right) compared with the exclusion curve from the HE-LHC projection studies in \protect\cite{CidVidal:2018eel} for $\sqrt{s}=27$ TeV at 15 ab$^{-1}$ of integrated luminosity.}
\label{figex2}
\end{figure}

The results are presented in  Figs.~(\ref{figex1}) and (\ref{figex2}) for the Run 2, HL-LHC and HE-LHC scenario respectively. The regions below the solid (violet) lines in Figs.~\ref{figex1}-\ref{figex2} indicate the parameter space for which 100\%, 95\% and 50\% of the events satisfy the unitarity bound, i.e. they define the regions where the model should not be trusted because unitarity is violated for such $(M, \Lambda)$ pairs. It is important to stress that the impact of the unitarity bound is strongly dependent on this fraction of events ($f_e$) that satisfy the condition in Eq.~(\ref{UNI_CI}). This also conforms with the results in~\cite{Endo:2014mja}, at least in the $(\Lambda, M)$ region considered here.
Moreover, we observe that there is a value of the compositeness scale $\Lambda$, which depends on the parton collision energy $\sqrt{\hat{s}}$ and the excited fermion mass $M$, above which the unitarity bound saturates. One can estimate an upper bound by setting the collision energy $\sqrt{\hat{s}}=\sqrt{s}$; it is represented with the dotted (black) line in Figs.~\ref{figex1} and \ref{figex2} for the corresponding nominal energies $\sqrt{s}=13, \, 14, \, 27$ TeV.
An approximated (maximal) value of $\Lambda \approx\sqrt{s/3}$ is obtained when $s \gg M^2 $.

The experimental collaborations quote routinely the largest excluded excited-state mass by intersecting the 95\% C.L. exclusion curves with the $M \leq \Lambda$ constraint (dot-dashed gray line and gray shaded region in Fig.~\ref{figex1} and \ref{figex2}). This is the widely adopted condition imposed on the model validity and it originates from asking the heavy excited states to be at most as heavy as the new physics scale $\Lambda$. Despite it is a reasonable constraint, it does not take into account the typical energy scale that enters the production process, i.e.~$\sqrt{\hat{s}}$. Comparing the unitarity bounds, as represented by the solid violet lines in Figure~\ref{figex1} and \ref{figex2} with the standard prescription $M \leq \Lambda$, one can frame the interplay between the two. On the one hand, for small values of the heavy neutrino masses, the unitarity bound is more restrictive than $M \leq \Lambda$ (at least for $f_e=100\%$ and $95\%$). On the other hand, for higher values of the masses the two becomes compelling. The relative importance depends on both the collider nominal energy and event fraction $f_e$.  

If one applies the unitarity bound to the experimental results by following the same prescription as outlined before for $M \leq \Lambda$, then the maximal neutrino mass values change accordingly. For example, for LHC Run 2, we find $M \le  3.6$ TeV for $\Lambda= 6.4 $ TeV instead of $M \le  4.6$ TeV ($\Lambda=4.6$ TeV), when the unitarity bound is required to be satisfied by the 100\% of events. The new constraint set by the unitarity bound offers an alternative theoretical and phenomenological guidance  for  ongoing and future experimental analyses on this effective composite model.

\section{Conclusions and outlook}
We studied the perturbative unitarity bound for the dimension-5 and dimension-6 gauge and contact Lagrangians for a composite-fermion model. It is well-know that an effective theory is valid up to energy/momentum scales smaller than the large energy scale that sets the operator expansion. Since collider experiments are delivering very energetic particle collisions, the applicability of effective operators can be put in jeopardy. In order to address this issue and to be on the safe side, one can impose the unitarity condition on the EFT parameters $(M, \Lambda, g^*, g)$ and the energy involved in a given process. To the best of our knowledge, such a constraint was not derived for the model Lagrangians in Eq.~(\ref{CI_Lag_LH}) and (\ref{GI_Lag_LH}). We have obtained the corresponding unitarity bounds  in Eqs.~(\ref{UNI_CI}) and (\ref{UNI_GI}), and presented them at this workshop. Thus, the applicability of the effective operators describing the production of composite neutrinos, and other excited states, has to be reconsidered accordingly. We have estimated a theoretical error on the unitarity bound, which is derived at leading order in the EFT expansion, by allowing up to 50\% of the events to evade the constraint. However, the 100\% line is the one that allows to avoid pathological points of the parameter space. At the end of the day, the experimental analyses rely on the leading order dimension-6 operators for the production of the massive excited fermions and no additinal term in the EFT expansion is considered. Therefore, the bound at 100\% is the one that should be considered to use the model in its region of validity especially because one assumes rather large couplings ($g_*^2=4\pi$) in the contact interactions scheme. 

\bibliographystyle{hieeetr}
\bibliography{unitarity.bib}

\end{document}